\documentclass[twocolumn,aps,showpacs,prx]{revtex4-1}

\usepackage{amsmath,amssymb,graphics,epsfig,epstopdf,color,multirow,array,verbatim,ulem,braket,tabularx}
\usepackage[colorlinks,linkcolor=blue,citecolor=blue,urlcolor=blue]{hyperref}
\usepackage{color}
\usepackage{graphicx}
\usepackage{epstopdf}
\usepackage{amsmath}
\usepackage{multirow}
\usepackage{ulem}
\usepackage{orcidlink}

\begin{document}

\title{Tunable surface electron gas and effect of phonons in Sr$_2$CuO$_3$: A first-principles study}

\author{Xin Du\orcidlink{0000-0003-1918-7568}$^{1,2}$}
\author{Hui-Hui He\orcidlink{0009-0000-7271-2443}$^{1,2}$}
\author{Xiao-Xiao Man\orcidlink{0000-0002-5172-3554}$^{1,2}$}
\author{Zhong-Yi Lu\orcidlink{0000-0001-8866-3180}$^{1,2}$}\email{zlu@ruc.edu.cn}
\author{Kai Liu\orcidlink{0000-0001-6216-333X}$^{1,2}$}\email{kliu@ruc.edu.cn}

\affiliation{$^1$School of Physics and Beijing Key Laboratory of Opto-electronic Functional Materials $\&$ Micro-nano Devices, Renmin University of China, Beijing 100872, China \\
$^2$Key Laboratory of Quantum State Construction and Manipulation (Ministry of Education), Renmin University of China, Beijing 100872, China}

\date{\today}

\begin{abstract}
While the conducting CuO$_2$ planes in cuprate superconductors have been widely recognized as a crucial component in producing high superconducting $T_\text{c}$, recent experimental and theoretical studies on Ba$_{2-x}$Sr$_x$CuO$_{3+}$$_\delta$ have also drawn much attention to the importance of Cu-O chains in one-dimensional (1D) cuprates. To better understand the cuprates containing Cu-O chains, here we have studied the electronic, magnetic, and phonon properties of Sr$_2$CuO$_3$ bulk and films based on the spin-polarized density functional theory calculations. We first reproduced the typical Mott insulator feature of the cuprate parent compound for bulk Sr$_2$CuO$_3$, and then built a Sr$_2$CuO$_3$ thin film with Cu-O chains exposed on the surface to directly investigate their characteristics. Different from the insulating bulk phase, the Sr$_2$CuO$_3$ surface shows interesting metallic properties. Further electronic structure calculations reveal the existence of spin-polarized electron gas between surface Sr atoms that strongly depends on the interchain coupling of Cu spins. Moreover, the phonon modes that involve the vibrations of in-chain and out-of-chain O atoms can induce strong charge and spin fluctuations in the surface layer of Sr$_2$CuO$_3$ film, which suggests significant multiple degree-of-freedom couplings that may be important for the superconductivity in 1D cuprates. Our work provides a comprehensive viewpoint of the properties of Cu-O chains in Sr$_2$CuO$_3$, facilitating a complete understanding of 1D cuprate superconductors.
\end{abstract}


\maketitle

\section{Introduction}
Cuprates, serving as the superconducting family with the highest superconducting transition temperature ($T_\text{c}$) at ambient pressure \cite{Marezio93, Antipov05, Tokura15}, have been one of the most important topics in condensed matter physics over the past decades \cite{Zaanen15, Shen03, Plakida10, Xue21, Hwang19, Shen21}. Despite extensive studies conducted to explore the conducting two-dimensional (2D) CuO$_{2}$ planes, which are generally accepted to play an active role in high $T_\text{c}$ \cite{Pickett89, Leggett06}, the origin of high-temperature superconductivity still remains a puzzle. On the other hand, in the history of superconductors, plenty of quasi-one-dimensional (1D) superconductors have also been reported \cite{Panagopoulos16, Zaikin08, Bao15, Sheng12, Wang19, Takeda93}. Therefore, in addition to the cuprates with 2D CuO$_{2}$ planes, the cuprates with the simple basic unit of 1D Cu-O chains may provide another opportunity to explore the microscopic mechanism of unconventional superconductivity.

Several alkaline-earth cuprates without intact 2D CuO$_{2}$ planes have manifested superconductivity \cite{Takeda93, Mueller20, Li19}. Early high-pressure experiments reported the synthesis of Sr$_{n+1}$Cu$_n$O$_{2n+1+}$$_\delta$ compounds and refined Sr$_{2}$CuO$_{3.1}$ ($n$ = 1, $T_\text{c}$ $\sim$ 70 K) with a K$_{2}$NiF$_{4}$ type structure \cite{Takeda93}. Subsequent neutron powder diffraction and electron diffraction measurements suggested that oxygen vacancies reside mainly on the CuO$_{2}$ planes \cite{Jorgensen94, Payne95}. Later on, the Cu K-edge x-ray absorption fine structure (XAFS) studies revealed that the structure of highly overdoped superconductor Sr$_{2}$CuO$_{3.3}$ ($T_\text{c}$ $\sim$ 95 K) does not contain complete CuO$_{2}$ planes \cite{Mueller20, Conradson20}. Similarly, our computational study \cite{Xiang19} on a newly synthesized cuprate superconductor Ba$_{2}$CuO$_{3+}$$_\delta$ \cite{Li19} also suggested that oxygen vacancies prefer to reside in the planar sites rather than the apical sites, emphasizing the Cu-O chain structure in Ba$_{2}$CuO$_{3+}$$_\delta$. Moreover, recent angle-resolved photoemission spectroscopy (ARPES) experiments on Ba$_{2-x}$Sr$_x$CuO$_{3+}$$_\delta$ \cite{Chen21} found an additional strong near-neighbor attraction, which may be attributed to the electron-phonon ($el$-$ph$) interaction, and the subsequent theoretical modeling work highlighted the effect of the O phonon modes in the Cu-O chains \cite{Devereaux21}. The above studies imply that the 1D Cu-O chains, beyond the well-known 2D CuO$_{2}$ planes, may play an unexpected role in some alkaline-earth cuprates with high superconducting $T_\text{c}$. Among the alkaline-earth cuprates with perfect 1D Cu-O chains, Sr$_{2}$CuO$_{3}$ has been intensively investigated as an ideal 1D Heisenberg $S$ = 1/2 antiferromagnetic (AFM) system \cite{Erwin95, Chou01, Schlappa12, Schlappa18}. Nevertheless, there is much less research on the couplings between multiple degrees of freedom in Sr$_2$CuO$_3$, which may be indispensable for a complete understanding of the exotic physical properties of 1D cuprates.

In this work, we have performed first-principles calculations to explore the lattice dynamics, electronic structure, as well as magnetic and phonon properties of Sr$_2$CuO$_3$ bulk and films. After checking the Mott insulating behavior of bulk phase, we studied the electronic structure and magnetism of Sr$_2$CuO$_3$ film with Cu-O chains exposed on the surface. Notably, we find the spin-polarized electron gas between surface Sr atoms depending on the interchain coupling of Cu spins. Furthermore, we investigated the effects of phonon and surface modification on the electronic and magnetic properties of surface Cu-O chains. Our findings about the tunable spin-polarized electron gas on surface and multiple-degree-of-freedom couplings of Sr$_2$CuO$_3$ film call for future experimental examination.

\section{Computational details}

The structural, electronic, and magnetic properties of Sr$_2$CuO$_3$ bulk and films were studied by using the spin-polarized density functional theory (DFT) calculations \cite{Kohn64,Sham65} with the generalized gradient approximation (GGA) of the Perdew-Burke-Ernzerhof (PBE) functional \cite{Ernzerhof96}, as implemented in the Vienna \textit{ab initio} simulation package (VASP) \cite{Kresse96}. The projector augmented wave (PAW) \cite{Blochl94} potentials with valence electron configurations of 4$s^2$4$p^6$5$s^2$, 3$d^{10}$4$s^1$, and 2$s^2$2$p^4$ were adopted for Sr, Cu, and O atoms, respectively. The kinetic energy cutoff of the plane wave basis was set to 520 eV. Monkhorst-Pack {\bf k}-meshes \cite{Pack76} of 8$\times$8$\times$4 and 4$\times$6$\times$1 were used for the $\sqrt{2}$$\times$$\sqrt{2}$$\times$1 bulk supercell and the thin films, respectively. All atomic positions were fully relaxed until the forces on atoms were smaller than 0.01 eV/{\AA} and the energy convergence criterion was set to 10$^{-6}$ eV. For the 7-layer Sr$_2$CuO$_3$ slab, the vacuum layer was set to 15 {\AA} in order to eliminate the artificial interactions among the image slabs along the (001) direction. Actually, we tested 20-{\AA} and 25-{\AA} vacuum layers to ensure that the vacuum layer of 15 {\AA} is thick enough to achieve good convergence. The effective Hubbard $U$ of 6.5 eV, which has been applied to the isostructural compound Ba$_2$CuO$_3$ \cite{Xiang19}, was utilized to describe the strong correlation effect among Cu 3$d$ electrons throughout all calculations. With $U_{\rm eff}$ = 6.5 eV, the lattice constants and band gap of the AFM ground state as well as the magnetic exchange strength (Tables S1, S2, and S3 in the Supporting Information (SI) \cite{SI}) are in good accordance with the experimental values \cite{Erwin95, Tokura96, Uchida96, Fujimori98}. The influence of the van der Waals interaction in the bare and I-absorbed Sr$_2$CuO$_3$ surfaces (films) was examined with the DFT-D2 method \cite{Scoles01, Grimme06}. Here, the spin-exchange interaction was determined based on an effective Heisenberg model \cite{Xaing08, Xaing09, Xaing16}: $H = J_{\bot} \sum_{<ij>} \Vec{S_i} \cdot \Vec{S_j} + J_l \sum_{<<ij>>} \Vec{S_i} \cdot \Vec{S_j}$, where $J_{\bot}$ and $J_l$ denote the respective couplings between the interchain and in-chain Cu spins, and $S$ is the local magnetic moment on Cu. The phonon properties were investigated via the frozen phonon approach \cite{Kawazoe97}. The polarization vector components $e_{\alpha i}^s$ ($\alpha$: atomic label; $i$ = $x$, $y$, $z$) of phonon modes $s$ were calculated through diagonalizing the dynamical matrix \cite{Gao05}. As introduced in our previous work on SrCuO$_2$ \cite{Liu24}, in a specific phonon mode $s$ with frequency $\omega_s$, the atoms could be displaced away from their equilibrium positions by $\pm\sqrt{2(n+1/2)\hbar/{m_\alpha\omega_s}}e_{\alpha i}^s$ ($n$ = 1 corresponds to a potential energy of 3$\hbar\omega_s$/2) along two opposite directions of the normal-mode coordinates. By this means, we could compare the physical properties of the structures with atomic displacements and those at equilibrium positions.

\section{Results and discussion}

\begin{figure}[!t]
\includegraphics[angle=0,scale=0.48]{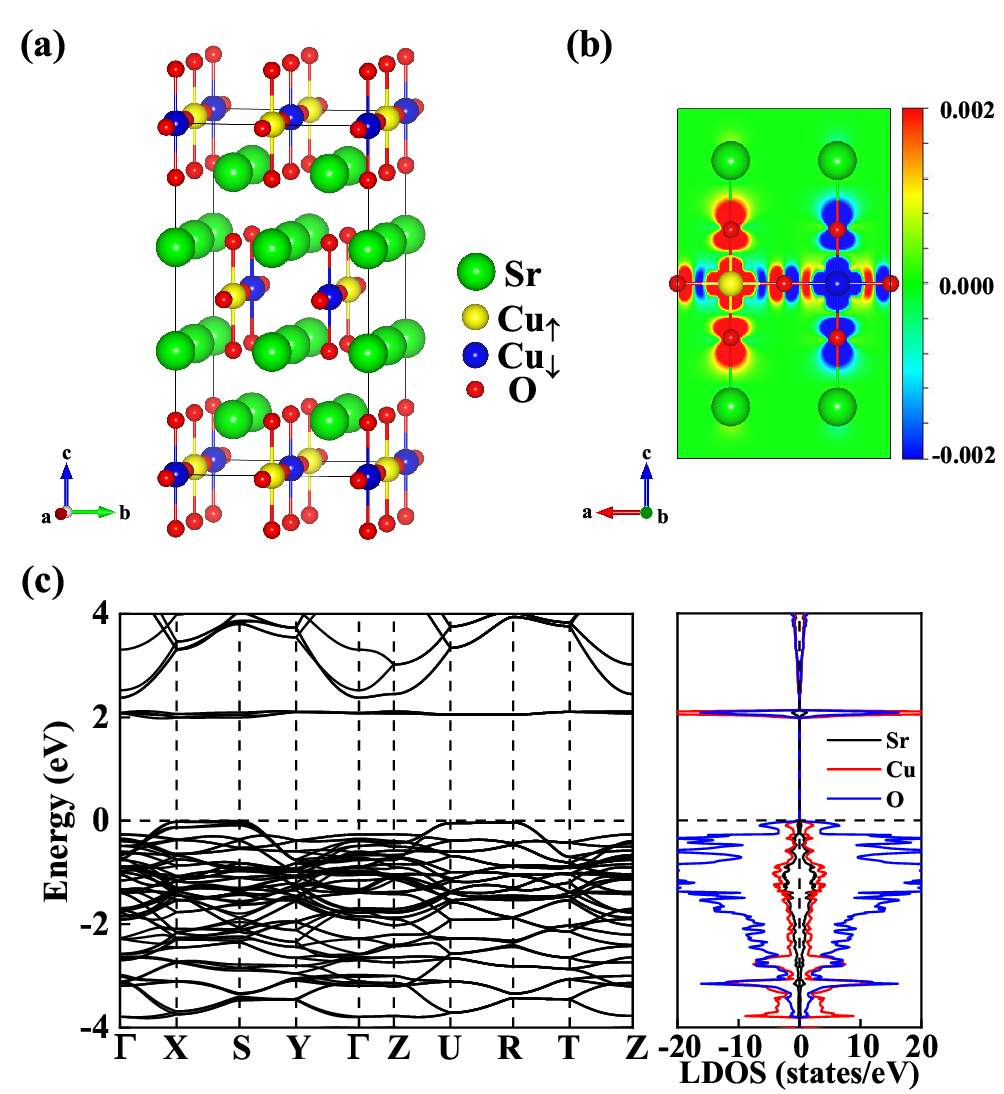}
\caption{(Color online) (a) Crystal structure, (b) spin density map, (c) electronic band structure and local density of states (LDOS) of bulk Sr$_2$CuO$_3$ in the AFM N\'{e}el state. The green and red balls represent Sr and O atoms, respectively. The yellow and blue balls denote the spin-up and spin-down Cu atoms, respectively. The two-dimensional (2D) spin density map is the (010) plane cut from the three-dimensional (3D) plot, and the color bar in (b) is in units of $e$/{\AA}$^3$.}
\label{fig1}
\end{figure}

\subsection{Crystal structure, magnetic configuration, and electronic properties of bulk Sr$_2$CuO$_3$}

Figures \ref{fig1}a and S1 of SI \cite{SI} show the crystal structure of Sr$_2$CuO$_3$, where the Sr-O, Cu-O, and Sr-O layers form a sandwich structure along the $c$ direction. Each Cu atom is coordinated with four O atoms, forming Cu-O chains along the $a$ direction. Because of the partially occupied Cu 3$d$ orbitals, the nonmagnetic (NM), ferromagnetic (FM), and three antiferromagnetic (AFM) states (N\'{e}el, AFM1, and stripe) were considered to study the magnetic configurations of Sr$_2$CuO$_3$. In the AFM N\'{e}el, AFM1, and stripe AFM states, the intrachain (interchain) couplings between Cu spins are AFM (AFM), AFM (FM), and FM (AFM), respectively (Fig. S1 of SI \cite{SI}). The total energies (with respect to that of the AFM N\'{e}el state) and the local magnetic moments on Cu atoms ($M_{\rm Cu}$) for these magnetic states are shown in Table S3 of SI \cite{SI}. The calculation results indicate that the AFM N\'{e}el state is the magnetic ground state, whose energy is 0.6, 137.7, 138.3, and 245.2 meV/Cu lower than those of the AFM1, stripe AFM, FM, and NM states, respectively. Due to the weak interchain magnetic coupling, the total energies and local moments $M_{\rm Cu}$ of the AFM N\'{e}el and AFM1 states, as well as those of the FM and stripe AFM states, are quite similar to each other.

We further explore the magnetic and electronic properties of Sr$_2$CuO$_3$ in the AFM N\'{e}el ground state (Fig. \ref{fig1}a). The spin density map plotted on the (010) plane clearly indicates that the Cu spins in each Cu-O chain are antiferromagnetically coupled along the $a$ axis and the O atoms are also spin-polarized by the neighboring Cu atoms (Fig. \ref{fig1}b). The calculated band structure and local density of states (LDOS) well reproduce the Mott insulator feature with a band gap of 1.98 eV, while the conduction band minimum (CBM) and valence band maximum (VBM) are contributed by Cu and O orbitals, respectively (Fig. \ref{fig1}c) \cite{Fujimori98}. According to previous studies on cuprates \cite{Xiang19, Olsen17, Reboredo14, Fujimori97}, various Hubbard $U_{\rm eff}$ values have been adopted to examine the above results of Sr$_2$CuO$_3$. Our calculations suggest that the chosen $U_{\rm eff}$ values, either 5.0, 6.5, 7.0, or 8.0 eV, would not change the relative stability among the magnetic states (Table S3 in the SI \cite{SI}). Notably, the band gap ($E_{\rm g}$ = 1.98 eV) of the AFM N\'{e}el state and the exchange coupling strength ($J_{\rm l}$ = 275 meV) calculated with $U_{\rm eff}$ = 6.5 eV (Tables S1 and S2 in the SI \cite{SI}) are in good agreement with the experimental values of 1.5 eV and 260 meV~\cite{Tokura96, Uchida96, Fujimori98}, respectively.

\subsection{Structural, magnetic, electronic and phonon properties of Sr$_2$CuO$_3$ thin film}

\begin{figure}[!t]
\includegraphics[angle=0,scale=0.54]{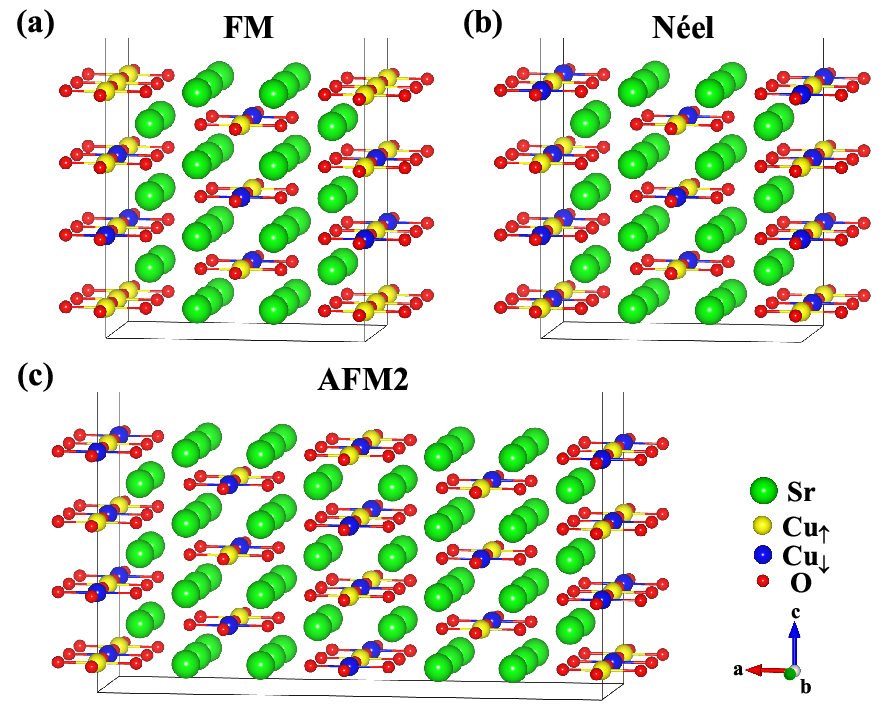}
\caption{(Color online) Crystal structure and typical magnetic configurations of the 7-layer Sr$_2$CuO$_3$ slab: (a) FM state, (b) AFM N\'{e}el state, and (c) AFM2 state. The green and red balls represent Sr and O atoms, respectively. The yellow and blue balls denote the spin-up and spin-down Cu atoms, respectively.}
\label{fig2}
\end{figure}

Our previous study on Ba$_{2}$CuO$_{3+}$$_\delta$ has proposed that the Cu-O chains may play an important role in its superconducting pairing \cite{Xiang19}. Recent experiments on the Ba$_{2-x}$Sr$_x$CuO$_{3+}$$_\delta$ superconductor have suggested that an additional strong interaction is attributed to the $el$-$ph$ coupling in Cu-O chains \cite{Chen21, Devereaux21}. We noticed that pristine Sr$_2$CuO$_3$ film with Cu-O chains exposed on the surface could be grown by orientation-controlled epitaxy \cite{Manako01}, and hence there is a chance to directly study the electronic, magnetic, and phonon properties of Cu-O chains on the surface of the Sr$_2$CuO$_3$ thin film.

\subsubsection{Crystal structure and magnetic configurations of Sr$_2$CuO$_3$ thin film}

We built a 7-layer Sr$_2$CuO$_3$ slab along the direction perpendicular to the Cu-O chains to simulate the Sr$_2$CuO$_3$ film (Fig. \ref{fig2}). Since the surface environment has changed in comparison with the bulk counterpart and may influence the magnetism of the top layer \cite{Lu18}, we have checked the magnetic configurations for the surface layer of Sr$_2$CuO$_3$ film. To determine the magnetic ground state, the NM and FM states of surface atomic layers containing Cu-O chains are considered, while other inner layers maintain the AFM N\'{e}el state of bulk Sr$_2$CuO$_3$ (Fig. \ref{fig2}a). Two AFM states (AFM N\'{e}el and AFM2 states) are also considered, in which the intrachain nearest-neighbor couplings between Cu spins are both AFM, while the interchain couplings are FM (Fig. \ref{fig2}b) and AFM (Fig. \ref{fig2}c), respectively. The total energy calculations show that the NM state is unstable and the AFM N\'{e}el state cannot be distinguished from the AFM2 state due to the weak interchain coupling. The energy of the AFM N\'{e}el state is only 12 meV/Cu lower than that of the FM state (Table S4 of SI \cite{SI}). In comparison with the bulk value of 138 meV/Cu, this indicates that the magnetic competition (fluctuation) is much stronger on the surface.

\begin{figure}[!t]
\includegraphics[angle=0,scale=0.54]{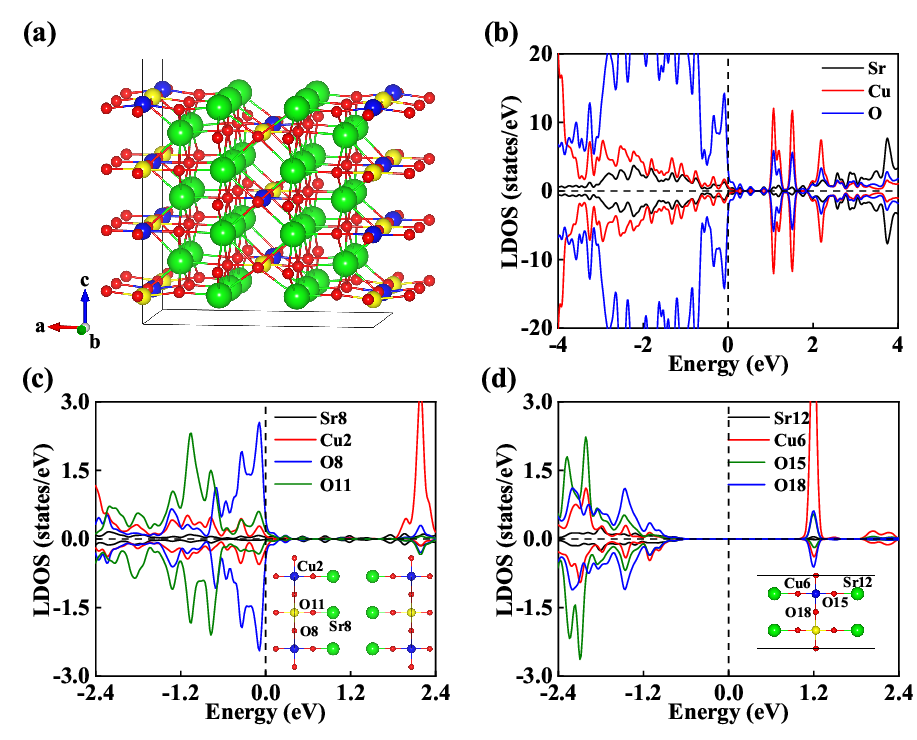}
\caption{(Color online) (a) Optimized crystal structure, (b) total, (c) surface-layer, and (d) inner-layer LDOS of Sr$_2$CuO$_3$ film in the AFM N\'{e}el ground state. The insets in (c) and (d) show the surface-layer and inner-layer structures, respectively.}
\label{fig3}
\end{figure}

\subsubsection{Electronic and magnetic properties of Sr$_2$CuO$_3$ thin film}

To explore the electronic properties of the Sr$_2$CuO$_3$ film in the AFM N\'{e}el ground state, the optimized crystal structure and LDOS are displayed in Figs. \ref{fig3} and S2 of SI \cite{SI}. Unexpectedly, unlike the insulator feature of bulk phase (bandgap of 1.98 eV) as shown in Fig. \ref{fig1}c, the LDOS of Sr$_2$CuO$_3$ film in the AFM N\'{e}el ground state exhibits metallic properties (Fig. \ref{fig3}b). The layer-resolved DOS calculations show that only the surface layers are metallic and the states around the Fermi level are mainly contributed by O atoms (Fig. \ref{fig3}c), while the inner layers retain the Mott insulator behavior with a bandgap of $\sim$ 1.62 eV (Fig. \ref{fig3}d). To better understand the origin of the metallicity, we analyzed the optimized crystal structure in detail (Fig. \ref{fig3}a). Although the chemical ratio of the surface layer is Sr:Cu:O = 2:1:3 as the bulk one, the lack of atoms on the vacuum side can cause the atomic distortion and the self-doping via the band bending of the surface layer, which lead to the metallic property. In addition, the results of the AFM2 state are quite similar to those of the AFM N\'{e}el state, and here we would not illustrate further (Fig. S2 of SI \cite{SI}).

\begin{figure}[!t]
\includegraphics[angle=0,scale=0.24]{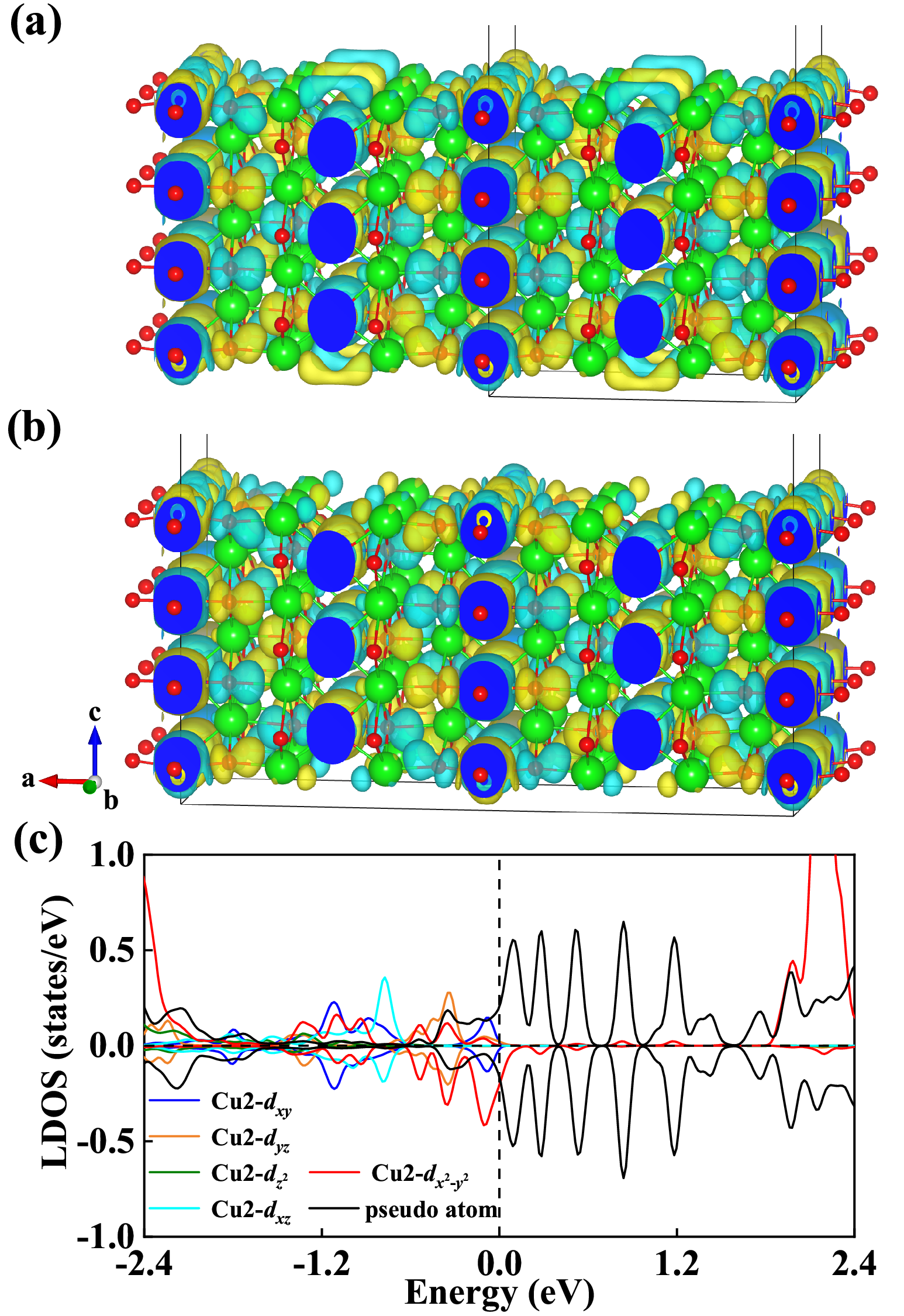}
\caption{(Color online) Spin density maps of Sr$_2$CuO$_3$ films in the (a) AFM N\'{e}el and (b) AFM2 states with the isosurface value of 0.0003 $e$/{\AA}$^3$. (c) The LDOS of pseudo atom (locating between surface Sr atoms) and surface Cu2 atom [Fig. \ref{fig3}(c)] in the AFM N\'{e}el state.}
\label{fig4}
\end{figure}

\begin{figure*}[!t]
\includegraphics[angle=0,scale=0.65]{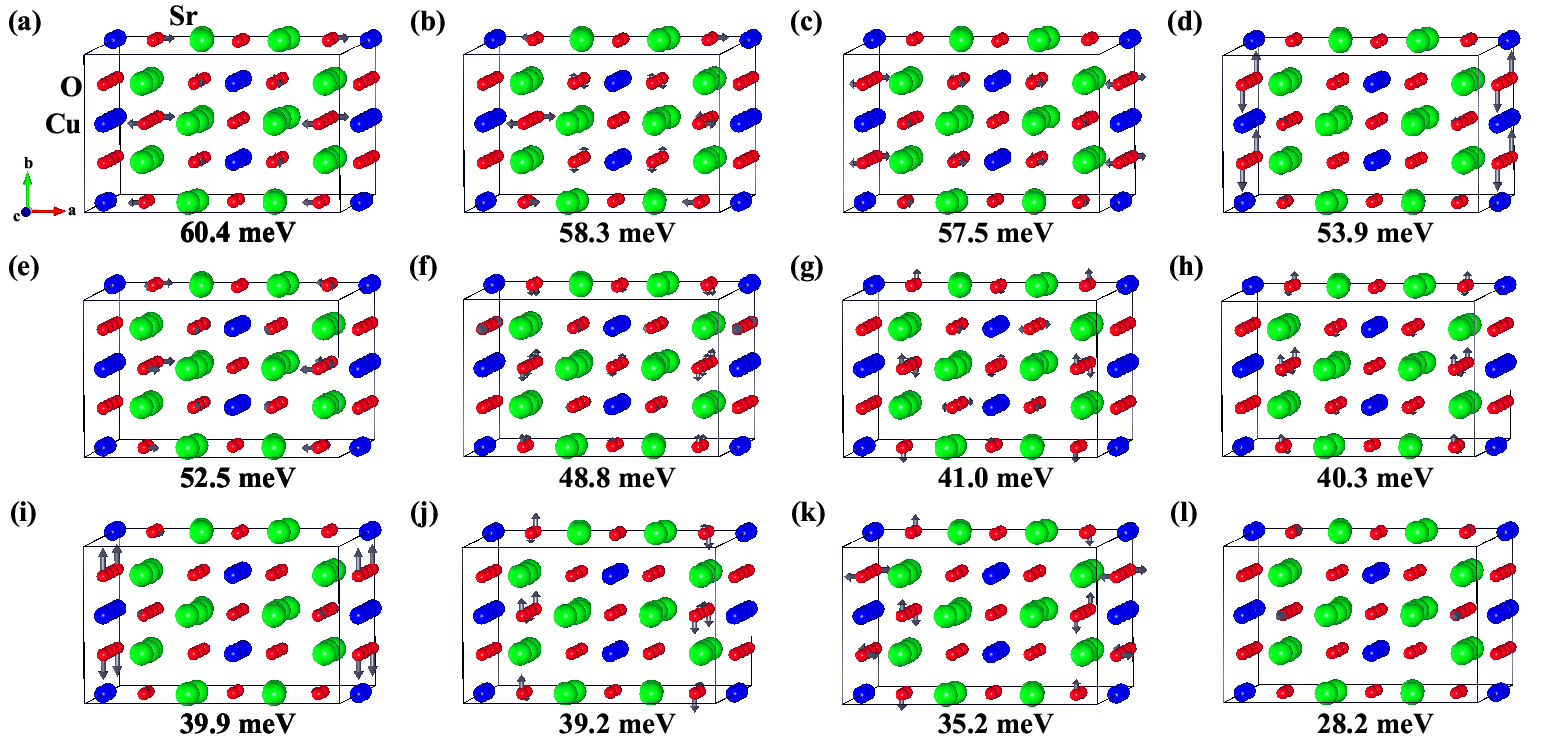}
\caption{(Color online) Top views of atomic displacement patterns for twelve optical phonon modes involving surface O atoms of Sr$_2$CuO$_3$ in the AFM N\'{e}el state. The gray arrows denote the directions and amplitudes of the atomic vibrations. The corresponding frequencies of the phonon modes are labeled in each panel.}
\label{fig5}
\end{figure*}

Spin density maps of Sr$_2$CuO$_3$ film in the AFM N\'{e}el and AFM2 states are plotted in Figs. \ref{fig4}a, \ref{fig4}b, and S3 of SI \cite{SI}. Interestingly, in these two AFM states with similar total energy (Table S4 \cite{SI}) and electronic properties (Figs. \ref{fig3} and S2 of SI \cite{SI}), the weak interchain coupling of surface Cu spins has a large impact on the spin densities between surface Sr atoms. Specifically, for the AFM N\'{e}el state with FM interchain coupling, there are spin-polarized electron gases between two rows of surface Sr atoms, which show alternating spin-up and spin-down polarizations along the $b$ axis (Figs. \ref{fig4}a, S3a, and S3c \cite{SI}). As for the AFM2 state, where the interchain coupling is AFM, there is no spin-polarized electron gas between two rows of surface Sr atoms (Figs. \ref{fig4}b, S3b, and S3d \cite{SI}). Compared with the bare spin density between Sr atoms in the bulk phase (Fig. \ref{fig1}b), these results indicate that the metallic surface of Sr$_2$CuO$_3$ film may have strong spin fluctuations due to the subtle interchain coupling. In addition, if a weak magnetic field is applied, the interchain interaction will favor the FM coupling instead of the AFM one, and would thus induce the appearance of spin-polarized electron gas between surface Sr atoms, i. e. the magnetic field can control the spin-polarized electron gas. We further added a pseudo atom to quantitatively describe the spin-polarized gas between two rows of surface Sr atoms in the AFM N\'{e}el state. The local magnetic moment of the pseudo atom is about 0.03 $\mu$${\rm _B}$, which is an order of magnitude smaller than that of the Cu atom ($\sim$ 0.6 $\mu$${\rm _B}$). The LDOS calculations suggest that the pseudoatom and the 3$d$ orbitals of surface Cu2 atoms have comparable contributions around the Fermi level (Fig. \ref{fig4}c). Therefore, the intrinsic spin fluctuations and the controllable conductive spin-polarized electron gas make the surface of Sr$_2$CuO$_3$ (Fig. \ref{fig2}) an interesting platform for quantum state exploration.

\subsubsection{Phonon effect on the electronic and magnetic properties of Sr$_2$CuO$_3$ thin film}

\begin{figure*}[tbh]
\includegraphics[angle=0,scale=0.59]{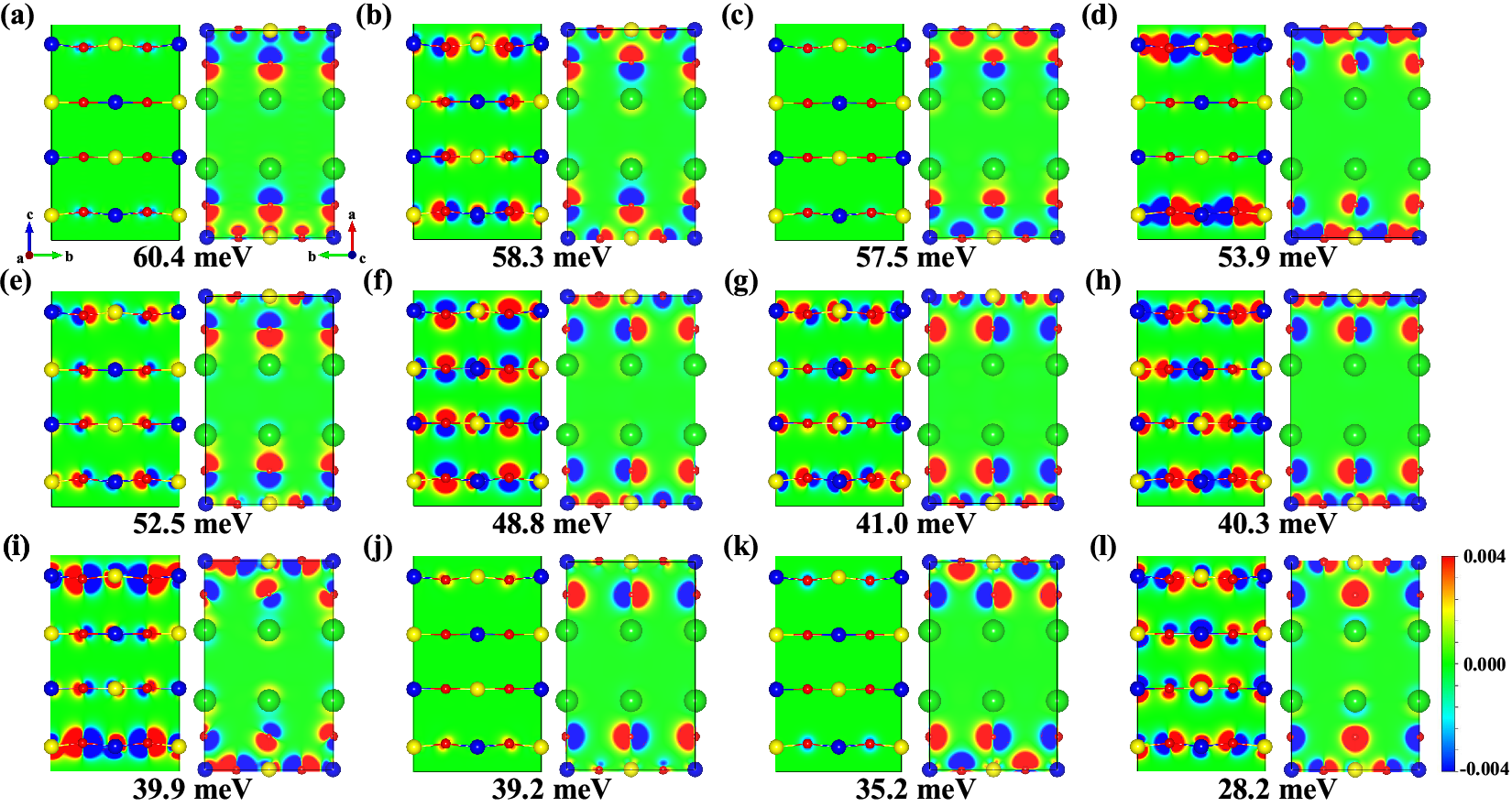}
\caption{(Color online) Differential charge density maps of the Sr$_2$CuO$_3$ slab in the AFM N\'{e}el state between the distorted structures induced by the atomic displacements of different O phonon modes (Fig. \ref{fig5}) and the equilibrium structure plotted on the (100) plane (side view) and the (001) plane (top view). The color bar is in units of $e$/{\AA}$^3$.}
\label{fig6}
\end{figure*}

\begin{figure*}[tbh]
\includegraphics[angle=0,scale=0.59]{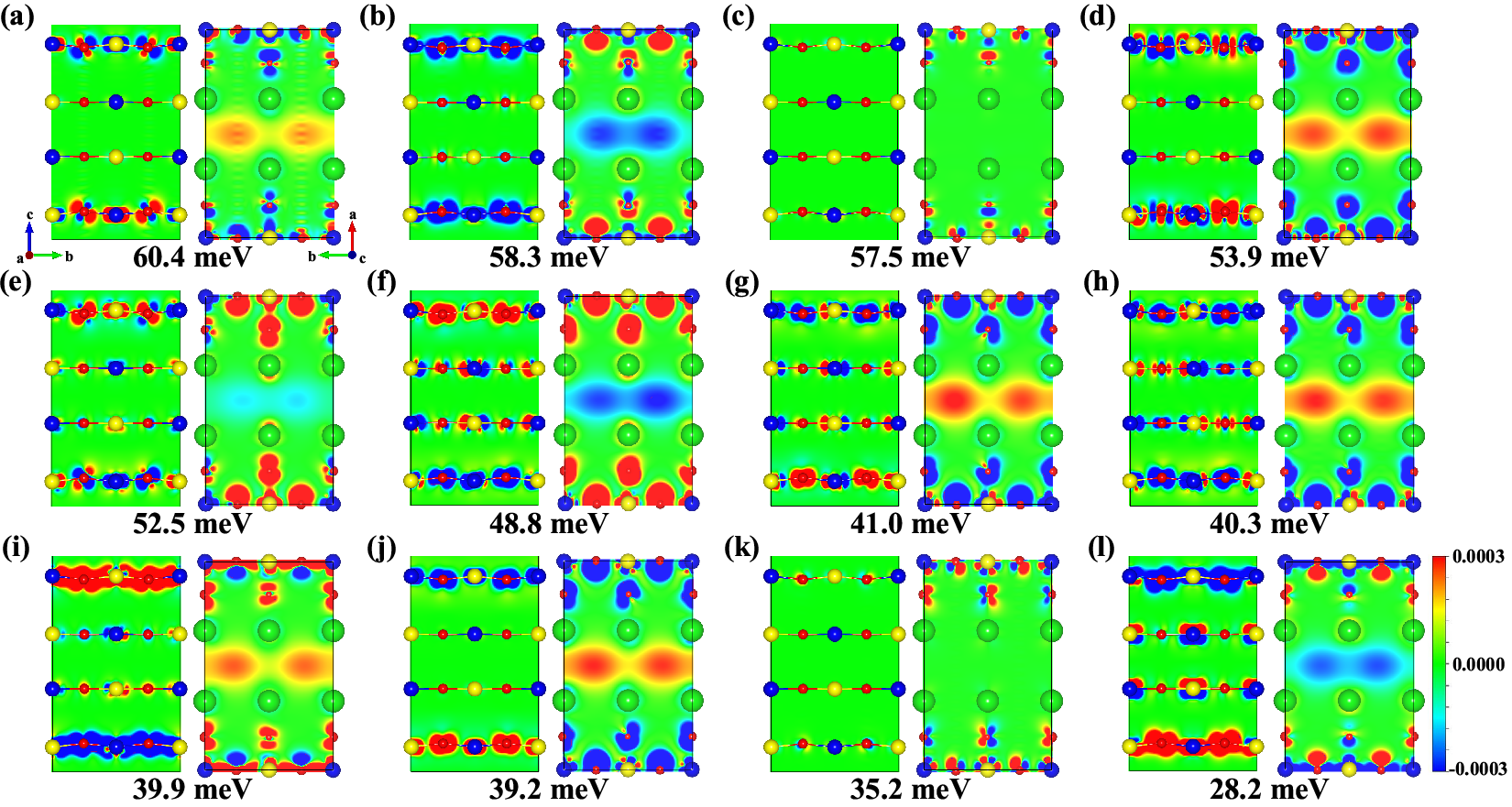}
\caption{(Color online) Differential spin density maps of the Sr$_2$CuO$_3$ slab in the AFM N\'{e}el state between the distorted structure induced by the atomic displacements of different O phonon modes (Fig. \ref{fig5}) and the equilibrium structure plotted on the (100) plane (side view) and the (001) plane (top view). The color bar is in units of $e$/{\AA}$^3$.}
\label{fig7}
\end{figure*}

Our recent study has revealed the notable role of O vibrations playing in the electronic and magnetic properties of the infinite-layer cuprate SrCuO$_2$ \cite{Liu24}; here we also focus on the phonon modes involving the surface O atoms in Sr$_2$CuO$_3$ film. For the 7-layer Sr$_2$CuO$_3$ slab, there are 84 atoms, corresponding to 252 phonon modes. The modes with considerable atomic displacements of surface O atoms are shown in Fig. \ref{fig5}, in which there are 4 and 8 phonon modes involving the vibrations of in-chain and out-of-chain O atoms, respectively. The in-chain O atoms have two vibration directions: along (Figs. \ref{fig5}d and \ref{fig5}i) and vertical to (Figs. \ref{fig5}c and \ref{fig5}k) the Cu-O chain (corresponding to the $b$ and $a$ axes, respectively). As for the out-of-chain (apical) O atoms, there are three vibration directions: along the $b$ (Figs. \ref{fig5}f, \ref{fig5}g, \ref{fig5}h, and \ref{fig5}j), $a$ (Figs. \ref{fig5}a, \ref{fig5}b, and \ref{fig5}e), and $c$ (Fig. \ref{fig5}l) axes. In addition, the vibrations of the O atoms in the same or opposite directions can further lead to different vibrational patterns. Specifically, the bond-stretching (Fig. \ref{fig5}i) and apical O (Fig. \ref{fig5}e) phonon modes have been investigated in previous theoretical studies, which were proposed to be associated with the $el$-$ph$ coupling in 1D cuprates \cite{Devereaux21, Johnston23}. There are several phonon modes similar to the two above modes: the mode with in-chain O atoms vibrating along the same direction (Fig. \ref{fig5}d, compared with Fig. \ref{fig5}i) and the modes with neighboring apical O atoms away from or close to the Cu-O chains (Figs. \ref{fig5}a and \ref{fig5}b, compared with Fig. \ref{fig5}e). Besides those modes, other phonon modes with the out-of-chain O vibrating parallel to the Cu-O chain are also uncovered (Figs. \ref{fig5}f, \ref{fig5}g, \ref{fig5}h, and \ref{fig5}j). Among all above phonon modes considered, the vibrations of out-of-chain O atoms along the $a$ and $c$ axes have the highest and lowest frequencies, respectively.

In consideration of previous modeling studies on 1D cuprates that have discussed the importance of the bond-stretching and apical O phonons in the $el$-$ph$ interaction \cite{Devereaux21, Johnston23}, here we study the phonon effect on the electronic and magnetic properties for a better understanding of the Cu-O chains on the Sr$_2$CuO$_3$ surface. We displaced the atoms in each specific phonon mode $s$ away from the equilibrium position with the corresponding potential energy of 3$\hbar\omega$$_s$/2. To visualize the phonon effect, we plotted the real-space differential charge and spin density maps between the structures with atomic displacements and the one at equilibrium position (Figs. \ref{fig6}, \ref{fig7}, S4, and S5 in the SI \cite{SI}). Interestingly, we found that all the phonon modes involving the atomic displacements of O atoms in Fig. \ref{fig5} can cause significant charge and magnetic fluctuations around the O and Cu atoms on the surface. To be specific, the change of charge density around the O atoms is very obvious; meanwhile the centers of the positive and negative charges do not coincide, indicating the existence of dynamical electrical dipoles on O atoms (Figs. \ref{fig6} and S4 of SI \cite{SI}). To further analyze the charge variations, we calculated the LDOS as a result of the O displacements of above twelve phonon modes (Fig. S6 \cite{SI}). As expected, the changes in the density of states of O are largest around the Fermi level, which are consistent with the results in Fig. \ref{fig6}. Notably, we found that the in-chain O vibrations (Figs. \ref{fig5}d and \ref{fig5}i) lead to stronger charge fluctuations (Figs. \ref{fig6}d and \ref{fig6}i) than those of other modes. On the other hand, we notice that there are magnetic fluctuations around Cu and O as well as spin-polarized electron gas between two rows of surface Sr atoms (Figs. \ref{fig7} and S5 of SI \cite{SI}), where the bond-stretching O phonon mode seems to cause the relatively largest magnetic fluctuations (Figs. \ref{fig5}i and \ref{fig7}i). Here, we attempt to understand why the bond-stretching O phonon (Fig. \ref{fig5}i) causes the most obvious charge and spin fluctuations (Figs. \ref{fig6}i and \ref{fig7}i). There may be two reasons: One is that the vibrations of in-chain O atoms can induce large charge redistributions between Cu and O atoms, thus influencing the charge and spin densities; the other is that according to the displacement formula in the section of COMPUTATIONAL DETAILS, the low frequency of this phonon mode (39.9 meV) corresponds to a large vibrational amplitude, which can directly affect the strength of fluctuations. Based on the above results, besides previous theoretical work that discussed one certain phonon mode contributing to the $el$-$ph$ interaction \cite{Devereaux21, Johnston23}, our results show that more phonon modes involving the vibrations of in-chain and out-of-chain O atoms can induce strong charge and magnetic fluctuations, which may be crucial to understand the properties of 1D cuprates.

\subsection{Modulating the surface properties of Sr$_2$CuO$_3$ film by iodine adsorption}

\begin{figure}[!t]
\includegraphics[angle=0,scale=0.32]{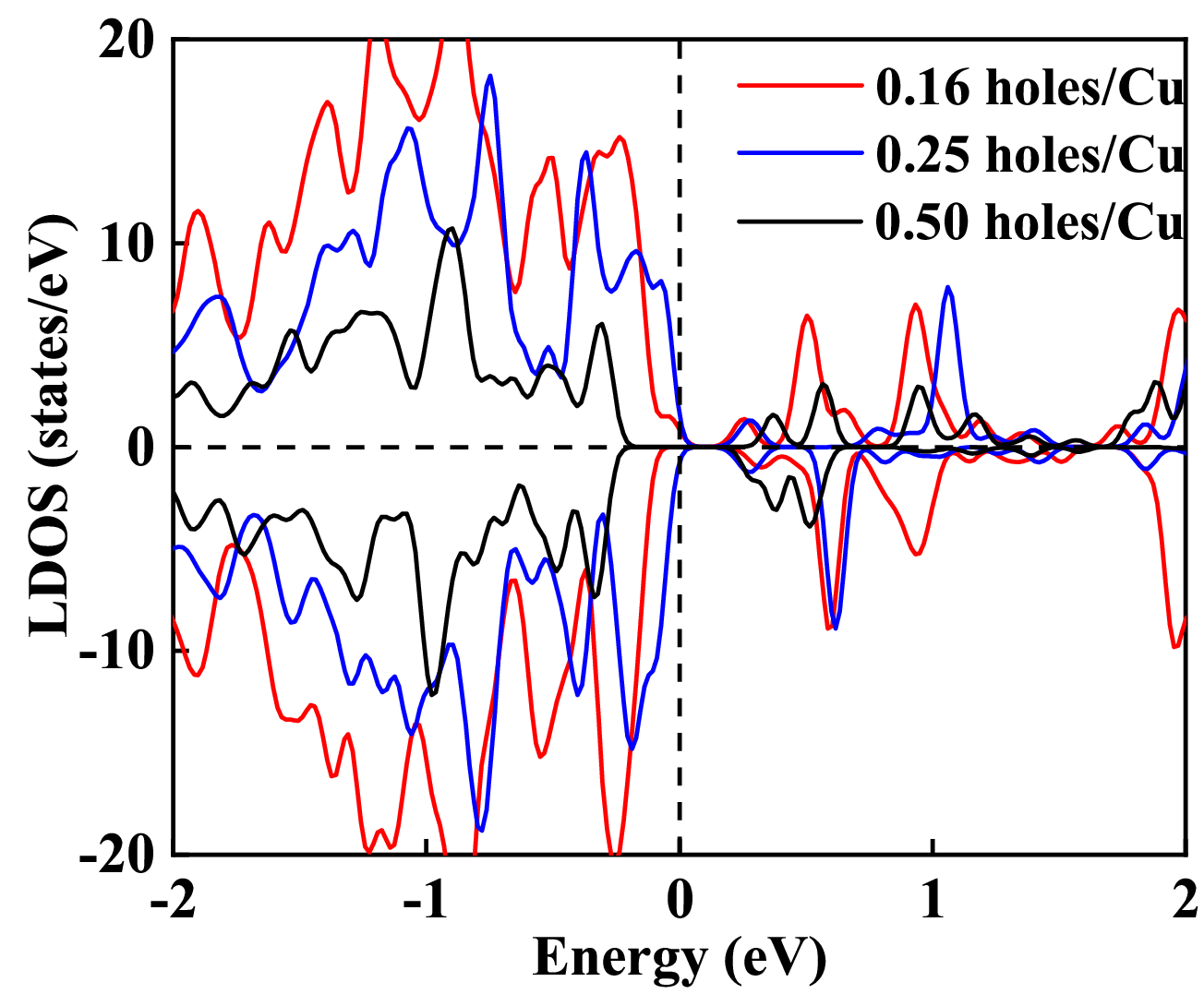}
\caption{(Color online) The LDOS for the top surface layer of I-adsorbed Sr$_2$CuO$_3$ films at the doping levels of 0.16, 0.25, and 0.50 holes per surface Cu atom.}
\label{fig8}
\end{figure}

In view of the important role of hole doping in cuprate superconductors, we next investigate the adsorption of the iodine (I) atom on the surface of the Sr$_2$CuO$_3$ film in order to modulate its electronic properties. As shown in Fig. S7 of SI \cite{SI}, we considered ten typical adsorption sites for I atom on the Sr$_2$CuO$_3$ surface at the doping level of 0.5 holes per surface Cu atom. According to their relative energies, the most stable structure is that I atom locates directly above Cu14 (Figs. S7f, S8a, and S8d \cite{SI}). Then, based on this stable adsorption site of the I atom, the doping levels of 0.25 and 0.167 holes per surface Cu were simulated with the 1$\times$2$\times$1 and 1$\times$3$\times$1 supercells, respectively (Fig. S8 \cite{SI}).

To study the magnetic and electronic properties of I-adsorbed Sr$_2$CuO$_3$ film, we determined the magnetic ground state under different doping levels (Table S5 of SI \cite{SI}) and calculated layered-resolved DOSs as in Figs. \ref{fig8} and S9 of SI \cite{SI}. Although the total DOSs all show metallic properties under different doping levels (Figs. S9a-c \cite{SI}), we find that the metallicity is mainly derived from the bottom surface layer (7th layer of the slab) (Figs. S9g-i \cite{SI}). Unexpectedly, with the increase of hole doping by I atom adsorption, the top surface layer (1st layer of the slab) gradually eliminates its metallicity, exhibiting diverse doped-related magnetic and electronic transitions from antiferromagnetic metal to ferrimagnetic metal and then to ferromagnetic semiconductor (Figs. \ref{fig8} and S9d-f of SI \cite{SI}). Here, we would like to address several points. First, although the DOS seems to be asymmetric at the doping of 0.16 holes/Cu (red line in Fig. \ref{fig8}), actually the Cu atoms are antiferromagnetically coupled and the whole system has a small net magnetic moment (Fig. S9f of SI \cite{SI}). Second, the stronger electronegativity of the I atom than that of the Sr atom leads to localized hole doping on the top surface, and therefore this surface cannot maintain the antiferromagnetic ground state at the doping of 0.25 holes per surface Cu (blue line in Fig. \ref{fig8} and Fig. S9e of SI \cite{SI}). Finally, the overdoped cuprates usually show ferromagnetic metal behavior with suppressed superconductivity \cite{Chakravarty07, Koike14, Greene20}; however, the I-adsorbed Sr$_2$CuO$_3$ surface at the doping of 0.5 holes per surface Cu is a ferromagnetic semiconductor (black line in Fig. \ref{fig8} and Fig. S9d of SI \cite{SI}). These results reflect the effective modulation of spin-polarized electron gas on the surface of Sr$_2$CuO$_3$ film by I adsorption.

\section{Discussion and Summary}

The structural, magnetic, electronic, and phonon properties of Sr$_2$CuO$_3$ bulk and films have been theoretically investigated. As expected, the bulk Sr$_2$CuO$_3$ exhibits the perfect Mott insulator feature (band gap of 1.98 eV) of the cuprate parent compound (Fig. \ref{fig1}c). Subsequently, we focus on the surface of the Sr$_2$CuO$_3$ film to study the Cu-O chains directly and have obtained the following results: (1) Unlike the bulk phase, the surface shows metallic properties, and the weak interchain coupling of Cu spins has a large impact on the spin-polarized electron gas between two rows of surface Sr atoms. (2) All phonon modes involving the vibrations of in-chain and out-of-chain O atoms can induce obvious charge and spin fluctuations, which may be crucial to a complete understanding of the superconductivity in 1D cuprates. (3) With the increase of hole doping, the I-adsorbed surface undergoes the transition from antiferromagnetic metal to ferrimagnetic metal and to ferromagnetic semiconductor.

Our computational results and theoretical analyses have identified the spin-polarized electron gas on the surface of the Sr$_2$CuO$_3$ film. Note that previous studies have revealed interesting 2D electron gas (2DEG) in the heterostructures. For example, the LaAlO$_3$/SrTiO$_3$ heterostructure has the superconducting 2DEG formed at the interface between two insulating oxides \cite{Reyren07}. It is also reported recently that a superconducting stripe state of 2DEG exists at the interface between insulating KTaO$_3$ and ferromagnetic insulator EuO \cite{Hua24}. Moreover, since the wide-band gap insulator SrTiO$_3$ is the most common substrate for oxide heterostructures, the 2DEG on the SrTiO$_3$ surface has also attracted much attention \cite{Fong22}. In fact, the vacuum can be regarded as an insulator, and our work on the Sr$_2$CuO$_3$ surface can also be viewed as a study of the interface between two insulators. Surprisingly, we find conductive and controllable spin-polarized electron gas that strongly depends on the coupling between interchain Cu spins. Moreover, we examined the evolution of the spin-polarized surface electron gas with the film thickness. For the 7-layer, 9-layer, and 11-layer Sr$_2$CuO$_3$ films, the contours of the spin-polarized surface electron gas look similar when the isosurface values of 3D plots are respectively set to 0.00030, 0.00028, and 0.00026 e/{\AA}$^3$ (Fig. S10 of \cite{SI}). This indicates that the spin-polarized surface electron gas decreases slightly with the increasing film thickness, but still remains on the thicker films. We believe that these unusual properties are important for heterostructure phenomena, spintronics application, and even for understanding cuprate superconductivity.

Additionally, we would like to emphasize that the study of Cu-O chains in Sr$_2$CuO$_3$ from a phonon perspective is indispensable in cuprate superconductors. Although the mechanism of unconventional superconductivity has not reached a consensus, a host of the inelastic neutron scattering (INS) and resonant inelastic x-ray scattering (RIXS) experiments have revealed that magnetic fluctuations in unconventional superconductors are much stronger than those in conventional metals, which suggests that magnetic fluctuations may play an important role in mediating the superconducting pairing \cite{Zaanen15, Tacon11, Dean13, Keimer00, Dogan01}. On the other hand, numerous ARPES and RXIS experiments as well as theoretical studies have confirmed that some specific phonon modes (e.g., the $B_{1}$$_g$ phonon of YBa$_2$Cu$_3$O$_{6+x}$) are associated with unconventional superconducting pairing \cite{Nagaosa02, Shen07, Nagaosa04, Zaccone23, Li20}. Here, our calculations on Sr$_2$CuO$_3$ not only show that multiple phonon modes can induce strong charge and magnetic fluctuations in Cu-O chains, but also suggest that phonons can effectively couple with other degrees of freedom. Notably, our previous work on infinite-layer SrCuO$_2$ (with 2D CuO$_2$ planes) has identified that only the full-breathing phonon mode, involving O in-plane vibrations along the Cu-O bonds, can cause significant fluctuations of local magnetic moments on O atoms and dramatic charge redistributions between Cu and O atoms \cite{Liu24}, which is different from present case of Sr$_2$CuO$_3$ with 1D Cu-O chains. Although we have not found a uniform law of phonon effect, it is clear that phonon modes in cuprate superconductors with different Cu-O structural units can induce significant charge and spin fluctuations. We draw attention to the important role that phonons play via the couplings with multiple degrees of freedom, which may provide insight to understanding the cuprate superconductivity.

To summarize, we have investigated structural, magnetic, electronic, and phonon properties of Sr$_2$CuO$_3$ bulk and films based on spin-polarized density functional theory calculations. The bulk phase shows typical Mott insulator features of the cuprate parent compound, while the surface of the Sr$_2$CuO$_3$ film shows interesting metallic properties. Through the direct study of Cu-O chains on the Sr$_2$CuO$_3$ surface, we find a spin-polarized electron gas between neighboring rows of surface Sr atoms dependent on the coupling of interchain Cu spins. And we find that multiple phonon modes, involving the vibrations of in-chain and out-of-chain O atoms, can induce obvious charge and spin fluctuations. Our work supplies a detailed microscopic picture of the electronic, magnetic, and phonon properties of cuprates containing Cu-O chains, which may facilitate the comprehensive understanding of transition metal oxides with low-dimensional structural units and even of the superconductivity in 1D cuprates.

\begin{acknowledgments}

This work was supported by the National Key R\&D Program of China (Grants No. 2022YFA1403103 and No. 2019YFA0308603), the Beijing Natural Science Foundation (Grant No. Z200005), and the National Natural Science Foundation of China (Grants No. 12174443, 11934020, and No. 11004243). K. L. was also supported by the National Key R\&D Program of China (Grant No. 2017YFA0302903). Computational resources have been provided by the Physical Laboratory of High Performance Computing at Renmin University of China and the Beijing Super Cloud Computing Center.

\end{acknowledgments}





\end{document}